\documentclass[smallextended,natbib,runningheads]{svjour3}
\smartqed  
\usepackage{graphicx}

\usepackage{amsmath,epsfig,epsf,psfrag}
\usepackage{amssymb}
\usepackage[latin1]{inputenc}
\usepackage{natbib} 
\usepackage{rotating}
\usepackage{amsbsy}
\usepackage{lscape}
\usepackage{color,latexsym,amsfonts,amscd}
\usepackage{array,tabularx,longtable,booktabs,threeparttable,colortbl,multirow,bigstrut}
\usepackage{longtable,supertabular}
\usepackage{mathrsfs}
\usepackage{txfonts}

\newcommand{\ben}{\begin{enumerate}}
	\newcommand{\een}{\end{enumerate}}
\newcommand{\be}{\begin{equation}}
\newcommand{\ee}{\end{equation}}
\newcommand{\bas}{\begin{eqnarray*}}
	\newcommand{\eas}{\end{eqnarray*}}
\newcommand{\ba}{\begin{eqnarray}}
\newcommand{\ea}{\end{eqnarray}}
\newcommand{\bit}{\begin{itemize}}
	\newcommand{\eit}{\end{itemize}}

\newcommand{\e}{ { \mathbb{E}}}

\newcommand{\bm}{\mbox{\boldmath $m$}}

\newcommand{\bgamma}{\mbox{\boldmath $\gamma$}}

\newcommand{\convergeto}{{\overset{d}{\longrightarrow}}}
\newcommand{\bvarphi}{  \bm{\varphi}   }
\newcommand{\bI}{{\bf I}}
\newcommand{\bzero}{{\bf 0}}
\newcommand{\bA}{{\bf A}}
\newcommand{\bE}{{\bf E}}
\newcommand{\bF}{{\bf F}}
\newcommand{\bV}{{\bf V}}
\newcommand{\bOmega}{{\bf \Omega}}

\newcommand{\pr}{ {\rm pr} }

\def\T{{ \mathrm{\scriptscriptstyle \top} }}

\begin{document}

\title{Empirical likelihood  meta analysis with publication bias correction under Copas-like selection model
}


\author{Mengke Li \and Yukun Liu \and
        Pengfei Li \and Jing Qin 
}

\authorrunning{Empirical likelihood meta analysis under Copas-like selection model} 

\institute{Mengke Li \and Yukun Liu  \at
              Key Laboratory of Advanced Theory and Application in Statistics and
              Data Science - MOE,
              School of Statistics, East China Normal University, Shanghai 200241, China \\
              \email{ykliu@sfs.ecnu.edu.cn}           
           \and
           Pengfei Li \at
              Department of Statistics and Actuarial Science,
              University of Waterloo, Waterloo, ON Canada N2L 3G1
           \and
           Jing Qin \at
            National Institute of Allergy and Infectious Diseases,
           National Institutes of Health, MD 20892, U.S.A.
}


\maketitle

\begin{abstract}
Meta analysis is commonly-used to  synthesize multiple results from individual
studies. However, its validation  is usually threatened by  publication
bias and between-study heterogeneity,   which can be  captured by the
 Copas selection model. Existing inference methods under this model
  are all based on conditional likelihood and  may not be fully
  efficient. In this paper, we propose a full  likelihood
  approach to meta analysis by integrating the conditional likelihood
  and a marginal semi-parametric empirical likelihood under a
  Copas-like selection model. We show that the maximum likelihood
  estimators (MLE)  of all the underlying parameters have a  jointly
   normal limiting distribution, and the full likelihood ratio follows
    an asymptotic central chisquare distribution. Our simulation results
    indicate that compared with  the conditional likelihood method,
    the proposed MLEs have smaller mean squared errors and the full
    likelihood ratio confidence intervals have more accurate coverage
    probabilities. A real data  example is analyzed to show the advantages
    of the full likelihood method over the conditional likelihood method.

\end{abstract}

\section{Introduction}
\label{intro}
Meta-analysis or systematic review  is the statistical technique of
collecting and synthesizing multiple published scientific results
from individual studies.  The most important  advantage of meta-analysis over
a single study is that  it usually has higher statistical power
and can answer research questions that cannot be answered
by a single study \citep{Jackson2011}.
Since formally introduced by \cite{Glass1976}
to evaluate the  effectiveness of psychological therapies,
it  has become  increasingly important and popular
in many fields of research, including medical, social, and
biological sciences  \citep{Egger2001,Cooper2009,Koricheva2012}.

However, as the basis of meta-analysis,
published scientific results may not be  representative of those from
all relevant studies, both published and unpublished,
due to the so-called publication bias, which is a well recognized threat to
the validation of the results of a meta analysis \citep{Rothstein2008,Jin2015}.
Typically, the studies used in the meta-analysis are biased towards
those which report statistically significant positive findings.
A standard meta-analysis may arrive at a misleading
conclusion that is biased toward significance or positivity \citep{Rosenthal1979} .

To retrieve valid inferences,  it is necessary and important to detect
and correct for publication bias  in meta analysis.
Many approaches have been developed in the literature
for this purpose. They can generally be divided into two categories.
The first category are  the funnel plot \citep{Light1984}
and related graphical methods \citep{Galbraith1988,Egger1997,Sterne2000,Sterne2001}. The funnel plot is a plot of
effect estimates from individual studies versus their precisions,
an asymmetric plot indicating potential publication bias.
The most well-known statistical test based on funnel plot asymmetry
is \cite{Duval2000a,Duval2000b}'s trim-and-fill test,
which is a rank-based data augmentation technique.
By formalizing the use of funnel plots, it estimates and adjusts
for the numbers and outcomes of missing studies.
However, simulation studies have found that the trim-and-fill
method may detect ``missing'' studies in a substantial proportion of
meta-analyses, even in the absence of bias.
In other words, it would add and adjust for non-existent studies
in response to funnel plot asymmetry arising from nothing more
than random variation \citep{Sterne2001}.

The second category are methods based on parametric models of the selection mechanism.
In these methods,  parametric distributions are imposed to characterize the underlying
publication mechanism by which effect estimates are selected to
be observed \citep{Rothstein2006} .
The most popular method is the  Copas selection model \citep{Copas1997,Copas1999,Copas2000, Copas2001},
which is derived  from the Heckman two-stage regression model \citep{Copas1997}.
Assuming  the most extreme studies are missing,
the trim-and-fill method often produces   excessively conservative inference \citep{Scharzer2010}.
By contrast,  the Copas model is more flexible, because
it   not only characterizes the heterogeneity and within-variation
of individual effect estimates, but also allows
the probability of selection  to depend on
both the effect estimate and its standard error.

Much attention has been paid to the Copas model  in recent years.
After analyzing 157 meta-analyses with binary outcomes, \cite{Carpenter2009}
concluded that the Copas selection model provided
a useful summary in 80\% of meta-analyses.
\cite{Scharzer2010}  demonstrated by  empirical evaluation indicated that
the Copas selection model is preferable to the trim-and-fill method
for selection bias in meta-analysis.
\cite{Mavridis2013} implemented the
Bayesian method for model fitting under the Copas model.
This method offers great flexibility  to incorporate in the model prior information
on the extent and strength of selection.
An unavoidable difficulty that blocks the wide application of the Copas selection model
is the frequent non-convergence in maximizing  the likelihood of the observed data.
A possible reason for this dilemma is that the data often contains very little information
about the underlying  parameters \citep{Copas2001}.
To overcome this problem, \cite{Ning2017}  re-casted
the biased-sampling problem as a missing data problem,
and proposed an EM  algorithm
to calculate the maximum likelihood estimators (MLE).

The main goal of meta-analysis is to estimate the overall effect size $\theta$ after adjusting
publication bias.  As an index of publication bias,
the number of unreported studies, $N_u$,  can also be important
to researchers \citep{Rosenthal1979,Fragkos2017}.
If one can estimate or determine $N_u$ in some way from the available data,
one can then compare this with one's knowledge about the field \citep{Gleser1996}.
To the best of our knowledge, the existing developments on the Copas selection model
are all based on the conditional likelihood which is the conditional joint distribution of data
given that they are observed or published.
Statistical inferences based on  conditional likelihood
are generally less efficient than those based on full likelihood
when we are estimating $\theta$.
Under the Copas selection model, the usual point estimator for $N = N_u+n$,
the total number of studies of interest,
is the inverse-probability-weighting estimator \citep{Mavridis2013},
whose asymptotical normality is used to construct Wald confidence intervals for $N$.
However such intervals may have poor coverage accuracy and
its lower bound can be even less than the sample size $n$,
which is clearly absurd \citep{Liu2017,Liu2018}.

In this paper,  we  formally show that  all the underlying parameters are
identifiable if publication bias exists  or equivalently the parameter $\rho$ is not equal to zero.
Assuming publication bias exists, we  focus on
the estimation of effect-size $\theta$ and the total number of studies, $N$.
Motivated the weaknesses of the conditional likelihood methods
under the Copas selection model,
we propose a full likelihood method for meta analysis
by integrating the conditional likelihood and a marginal semi-parametric likelihood.
We make the same model assumptions as \cite{Ning2017} ,
use \cite{Owen1990}'s empirical likelihood (EL) to handle the nonparametric
distribution of the standard errors of the individual effect sizes,
and finally derive the marginal semi-parametric  likelihood.
We show that the proposed MLEs of all the
underlying parameters have a jointly normal limiting  distribution,
and the full likelihood ratio follows an asymptotic central chisquare distribution.
In particular,  we propose to  construct confidence intervals for
the effect size and the total number of studies, and
test the existence of publication bias by
the corresponding full likelihood ratio tests.
Our simulation results indicate that the full maximum likelihood estimators
have smaller mean squared errors than
the conditional-likelihood-based estimators.
Also the full likelihood ratio confidence intervals for
the effect size and the total number of studies
have more accurate coverage probabilities than
the  Wald intervals under the conditional likelihood.

The paper proceeds as follows.
In Section 2, we introduce the Copas-like model of \cite{Ning2017},
present our semi-parametric full likelihood method,
and investigate the large-sample properties of the MLE and
the likelihood ratio test.
An algorithm to calculate the proposed MLEs are also provided.
Section 3 contains simulation results.
Section 4 is devoted to two real-life data analyses.
We end in Section 5 with some discussions.
For clarity, all proofs are postponed to the supplementary material.

\section{Full likelihood approach and its properties}

\subsection{Model set-up}

Let $N$ be the total number of studies of interest, including published and unpublished.
For study $i$, let $\theta_i^*$  denote the estimated
effect size and $s_i^{*}$  the estimated standard variance of $ \theta_i^*$.
We make the same assumptions on data as  \cite{Ning2017}.
Specifically, suppose that $(  \theta^*, s_i^*)$ ($1\leq i \leq N$) are
independent and identically distributed (IID), and that
$ \theta^*$ and $s_i^*$ are also independent of each other.
Given $s_i^{*}$,  \cite{DerSimonian1986}  modelled
$  \theta_i^*$  by a random effect  model
\begin{equation}
\label{rem}
\theta_i^*  = \theta + \tau u_i + s_i^* \epsilon_i,
\end{equation}
where the random effect $u_i$ and the error  $\epsilon_i$
are independent, and both follow the standard normal distribution.
Here $\theta$ is the underlying effect size and
$\tau^2$ describes the between-study
heterogeneity.
To characterize the publishing mechanism,
\cite{Copas1997}  proposed a separate selection model
that uses a latent variable
\begin{equation}
\label{lv}
Z_i  = \gamma_1+\gamma_2/s_i^* + \delta_i,
\end{equation}
where  $(\epsilon_i, \delta_i)$ are
independent and identically distributed
from a bivariate standard normal distribution
with correlation coefficient $\rho$.
Study $i$ is assumed to be published if  $Z_i>0$.
Suppose there are $n$ studies published  with estimated effect sizes
and estimated standard variances $\{ (\theta_1, s_1), \ldots, (\theta_n, s_n)\} $.
We wish to estimate  the effect size $\theta$,
heterogeneity $\tau$,  the total number $N$ of studies,
the marginal distribution  function $F(x)$  of $s_i^*$,
and  the marginal distribution $G(x)$ of $ \theta_i^*$, after adjusting for publication bias.

\begin{lemma}
	If $\rho\neq 0$,  the parameters $ \gamma_1, \gamma_2, \rho, \tau $ and $\theta$ are
	all identifiable.
\end{lemma}

Unlike \cite{Ning2017},  we assume that $\rho\neq 0$ in this paper,
therefore the observations $ (\theta_i, s_i)$'s constitute  a biased sample of
the study of interest, or equivalently publication bias exists.
Lemma 1 implies that  all the parameters are identifiable  in this case.

\subsection{Full semiparamtric  likelihood}

Let  $\pr$ denote the probability density/mass function
of a continuous/discrete random variable.
We assume that the observations $  (\theta_1, s_1), \ldots, (\theta_n, s_n) $
are independent  given  $n$, the number of observations.
It follows that the full likelihood is
\[
\tilde  L
= \pr\{ n, ( \theta_i, s_i), i=1, \ldots, n\}
= \pr(n)\times\prod_{i=1}^n  \pr  ( \theta_i, s_i).
\]
Without otherwise statement, we use the same notation
to denote both a random element and its realization,
whose meanings can be clear from the context.
For example,  $\pr(z)$  denotes
the  density function of a random element $z$ at $z$.

According to its generating process,
$(\theta_i, s_i)$ has the same distribution as $\{ (\theta_i^*, s_i^*)| Z_i>0\}$.
Thus formally  the full likelihood can be written as
\ba
\tilde  L
&=& \pr(n)\times \prod_{i=1}^n \pr  ( \theta_i^*=\theta_i, s_i^* = s_i | Z_i>0  ) \nonumber \\  %
&=& \pr(n)\times \prod_{i=1}^n \frac{ \pr(    Z_i>0 |  \theta_i^*=\theta_i, s_i^*=s_i )
	\pr(    \theta_i^*=\theta_i  | s_i^*=s_i ) \pr(s_i^*=s_i)}{ \pr(Z_i>0)}.
\label{conditional-probs}
\ea
Obviously,
\(
\pr(n) = { N \choose  n} \alpha^n (1-\alpha)^{N-n}
\)
with $\alpha = \pr(Z_i>0)$.
Lemma \ref{lemma-prob} presents the other  conditional probabilities and densities in
\eqref{conditional-probs}.

\begin{lemma}
	\label{lemma-prob}
	Let  $\phi(x)$ and $\Phi(x)$  denote
	the  standard normal density and distribution functions,
	and    $\bgamma = (\gamma_1, \gamma_2, \rho, \tau, \theta)^\T$.
	We have
	\ba
	\pr(Z_i>0| \theta_i^*=\theta_i,  s_i^*=s_i) &=& \Phi\{v_i(\bgamma)\},
	\label{lemma-prob-eq1} \\
	\pr( \theta_i^*=\theta_i| s_i^*=s_i)
	&=&
	\frac{1}{\sqrt{2\pi(\tau^2+s_i^2)}} \exp\left\{  - \frac{(  \theta_i - \theta)^2}{2(\tau^2+s_i^2)} \right\},
	\label{lemma-prob-eq2}
	\\
	\pr(Z_i>0| s_i^*=s_i) &=& \Phi(\gamma_1+\gamma_2/s_i ),
	\label{lemma-prob-eq3}
	\ea
	where
	\ba
	\label{vi-gamma}
	v_i(\bgamma )
	= \frac{ \gamma_1+ (\gamma_2/s_i) +  \rho s_i ( \theta_i - \theta)/( \tau^2 + s_i^{2} ) }{
		\sqrt{1- \rho ^2 s_i^{2}/(\tau^2 + s_i^{2})}  }.
	\ea
\end{lemma}

Lemma \ref{lemma-prob} implies that the full log-likelihood is
\bas
\tilde \ell
&=&
\log { N \choose  n}
+ (N-n) \log(1-\alpha)
+\sum_{i=1}^n \Big[ \log\{ \Phi(v_i(\bgamma)) \}
-\frac{1}{2} \log(\tau^2 + s_i^{2} )
-    \frac{ (\theta_i  - \theta)^2 }{2(\tau^2 + s_i^{2}) } \Big] \\
&&
+ \sum_{i=1}^n \log\{ \pr(s_i^*=s_i) \}.
\eas
We use  Owen (1988, 1990)'s EL method to handle the distribution function
of $s_i^*$.  Let $p_i=\pr(s_i^*=s_i)$.
Since
$
\pr(Z_i>0|s_i^*=s_i) = \Phi(\gamma_1+\gamma_2/ s_i)
$,
we have
\bas
\alpha = \pr(Z_i>0) = \e \{ \pr(Z_i>0|s_i^*=s_i)  \} =  \int  \Phi(\gamma_1+\gamma_2/s) \pr(s_i^*=s) ds.
\eas
Hence the feasible $p_i$'s satisfy
\ba
\label{constr}
p_i\geq 0, \quad
\sum_{i=1}^n p_i =1, \quad
\sum_{i=1}^n p_i  \Phi(\gamma_1+\gamma_2/s_i ) = \alpha.
\ea

With $p_i$ in place of $\pr(s_i^*=s_i)$,
the maximizer of $\tilde \ell$ with respect to $p_i$'s
under the constraints in \eqref{constr}  is
$
p_i = n^{-1} [1+\lambda\{ \Phi(\gamma_1+\gamma_2/s_i ) - \alpha \}]^{-1},
$
where $\lambda$ is the solution to
\ba
\label{eq-lambda}
\sum_{i=1}^n \frac{\Phi(\gamma_1+\gamma_2/s_i ) - \alpha}{1+\lambda\{ \Phi(\gamma_1+\gamma_2/s_i ) - \alpha \}} = 0.
\ea
Accordingly, we have the profile log EL  (up to a constant)
\bas
\ell (N, \alpha, \bgamma  )
&=&
\log { N \choose  n}
+ (N-n) \log(1-\alpha)
+\sum_{i=1}^n  \left[ \log\{ \Phi(v_i(\bgamma)) \}
-\frac{1}{2} \log(\tau^2 + s_i^{2} ) \right.\\
&&
- \left. \frac{ (\theta_i - \theta)^2 }{2(\tau^2 + s_i^{2}) } \right]
- \sum_{i=1}^n \log[ 1+\lambda\{ \Phi(\gamma_1+\gamma_2/s_i) - \alpha \} ].
\eas

\subsection{Estimation and Asymptotics}
We propose to estimate  $(N, \alpha, \bgamma)$ by the
MLEs
$
(\widehat N, \widehat \alpha ,\widehat \bgamma  )
= \arg\max \ell (N, \alpha,\bgamma ).
$
Define the likelihood ratio function to be
\(
R(N, \alpha,\bgamma  )
= 2\{ \ell  (\widehat N, \widehat \alpha , \widehat \bgamma  ) -   \ell (N, \alpha, \bgamma ) \}.
\)
This section  investigates the
asymptotical properties of the MLEs
and the likelihood ratio test statistic.

For ease of presentation,  we use $\bgamma_{12}$ and $\bgamma_{45}$ to denote $(\gamma_1, \gamma_2)^\T$
and $(\tau, \theta)^\T$, respectively, and define
\bas
f_{1}(\theta_i, s_i; \bgamma) &=& \pr(Z_i>0| \theta_i^*=\theta_i,  s_i^*=s_i) = \Phi\{v_i(\bgamma)\}, \\
f_{2}( \theta_i, s_i; \bgamma_{45}) &=& \pr( \theta_i^*=\theta_i| s_i^*=s_i)
=
\{2\pi(\tau^2+s_i^2) \}^{-\frac{1}{2}}  \exp\left\{  - \frac{(  \theta_i - \theta)^2}{2(\tau^2+s_i^2)} \right\}, \\
f_{3}(s_i; \bgamma_{12}) &=& \pr(Z_i>0| s_i^*=s_i) = \Phi(\gamma_1+\gamma_2/s_i ).
\eas
Let $(N_0, \alpha_0, \gamma_0)$ be the truth of $(N, \alpha, \bgamma)$
with $\bgamma_0 = (\gamma_{10}, \gamma_{20}, \rho_0, \tau_0, \theta_0)^\T$.
Throughout the paper, we   use  $\bgamma_{12,0}$ and $\bgamma_{45,0}$
to denote the   truths of $\bgamma_{12,0}$ and $\bgamma_{45,0}$, respectively.
Define  $\bA^{\otimes 2} = \bA \bA^\T $ for a matrix or vector $\bA$,
and $\bA^{\oplus 2} = \bA+\bA^\T$ for a square matrix $\bA$,
$\bE_{12} = (\bI_{2}, \bzero_{2\times 3})^\T$, and
$\bE_{45} = (\bzero_{2\times 3}, \bI_{2})^\T$ with
$\bI_{k}$  the $k\times k$ identity matrix.
We use $\nabla_{\bgamma}$ to denote the differentiation operator with respect to $\bgamma$.
Let
$
\varphi_1=\e [  \{   f_3(s_i^*; \bgamma_{12,0})   \}^{-1} ]
$,
$
\bvarphi_2=\e\left\{    \nabla_{\bgamma_{12}}  \log f_3(s_i^*; \bgamma_{12,0})
\right\},
$
$\bF_{1} = (\bI_{4}, \bzero_{4\times 3})$ and  $\bF_{2} = (\bzero_{5\times 2}, \bI_{5})$.
Define
\ba
\bOmega=\bF_2^\T \bV_{c}  \bF_2 + \bF_1^\T\tilde \bV_{m} \bF_1,
\label{Omega}
\ea
where
\bas
\bV_c
&=&
\e  \frac{  \{ \nabla_{\bgamma}   f_{1}(\theta_i^*, s_i^*; \bgamma_0)\}^{\otimes 2} }{  f_{1}(\theta_i^*, s_i^*; \bgamma_0)}
+
\e\Big[  \bE_{45}\int  \frac{ \{ \nabla_{\bgamma_{45}}   f_{2}( t, s_i^*; \bgamma_{45,0}) \}^{\otimes 2}}{
	f_{2}( t, s_i^*; \bgamma_{45,0}) }
f_{1}(t, s_i^*; \bgamma_0)   dt \bE_{45}^\T \Big]   \\
&&
+
\e\Big[  \int  \nabla_{\bgamma}   f_{1}(t, s_i^*; \bgamma_0)
\nabla_{\bgamma_{45}^\T}   f_{2}( t, s_i^*; \bgamma_{45,0})  dt \bE_{45}^\T
\Big]^{\oplus 2}
-
\e\Big[   \bE_{12} \frac{ \{ \nabla_{\bgamma_{12}} f_{3}(s_i^*; \bgamma_{12,0}) \}^{\otimes 2}   }{ f_{3}(s_i^*; \bgamma_{12,0}) }
\bE_{12}^\T\Big]
\eas
and
\bas
\tilde \bV_m
&=&
\left(
\begin{array}{ccc}
	\frac{\alpha_0}{1-\alpha_0} &  \frac{1}{1-\alpha_0} & \bzero   \\
	\frac{1}{1-\alpha_0}  &    \frac{ 1-\varphi_1  }{(1-\alpha_0)(1  - \alpha_0 \varphi_1) } &
	\frac{ \bvarphi_2^\T  }{1  - \alpha_0 \varphi_1 }  \\
	\bzero      &  \frac{ \bvarphi_2  }{1  - \alpha_0 \varphi_1 } &
	-
	\frac{\alpha_0 \bvarphi_2^{\otimes 2} }{1  - \alpha_0 \varphi_1 }
\end{array}
\right).
\eas

\begin{theorem}
	\label{asy-full-likelihood}
	
	Assume Conditions C1 and C2 in the supplementary materials, $\rho_0 \neq 0$,
	and that the matrix $\bOmega$  defined in \eqref{Omega}  is positive definite.
	As $N_0\rightarrow\infty$, the following results hold.
	\bit
	\item[(1)]
	$
	N_0^{1/2}(\widehat N/N_0 - 1, \;  \widehat \alpha-\alpha_0, (\widehat \bgamma - \bgamma_0)^{\T}  )
	\convergeto  N(\bzero, \;  \bOmega^{-1})
	$, where $\convergeto$ stands for convergence in distribution.
	\item[(2)]
	$
	N_0^{1/2}(\widehat N/N_0 - 1  )
	\convergeto  N(0, \;  \sigma^2),
	$
	and
	$
	N_0^{1/2}(\widehat \bgamma - \bgamma_0  )
	\convergeto  N(\bzero, \;  \bV^{-1} ),
	$
	where $\sigma^2$ is the $(1,1)$ element of $ \bOmega^{-1}$
	and $\bV^{-1}$ is the down-right $5\times 5$ submatrix of $\bOmega^{-1}$.
	\item[(3)]
	The likelihood ratio
	$
	R(N_0, \alpha_0, \bgamma_0)
	= 2 \{ \ell(\widehat N, \widehat \alpha, \widehat \bgamma) - \ell(N_0, \alpha_0, \bgamma_0) \}
	\convergeto \chi_7^2.
	$
	\eit
	
\end{theorem}

The proof of result (3) of Theorem \ref{asy-full-likelihood}  (See the supplementary material)
implies that the likelihood ratio statistic of testing any subvector of $(N, \alpha, \bgamma)$
also follows an asymptotic   central chisquare distribution.
This result can be used to construct likelihood ratio confidence intervals for
any of the parameters $\theta, \tau, N, \rho $ and  $\alpha$ with asymptotically correct coverage probabilities.

The proposed full likelihood method can conveniently
provide  consistent estimators for the marginal distributions
of $s_i^*$ and $\theta_i^*$.
Given   $\widehat \alpha$ and  $\widehat \gamma$,
the MLE of the distribution function $F(x)$ of $s_i^*$
is
\bas
\widehat F(s)=\sum_{i=1}^n \widehat p_i I( s_i \leq s)
=\dfrac{1}{n} \sum_{i=1}^n \dfrac{1}{1+\widehat\lambda\{ \Phi(\widehat\gamma_1+\widehat\gamma_2/s_i)-\widehat\alpha\}} I( s_i \leq s),
\eas
where  $\widehat\lambda$ is the solution to Equation \eqref{eq-lambda}
with  $(\gamma_1, \gamma_2, \alpha)$ replaced by $(\widehat \gamma_1, \widehat \gamma_2, \widehat \alpha)$.
To estimate the marginal distribution $G(t)$ of $\theta_i^*$,
we   rewrite  $G(t)$ as
\bas
G( t)
& =&  \int_{-\infty}^t  \int     \pr( \theta_i^*= r | s_i^*=s) dF(s) dr \\
& =&  \int_{-\infty}^t  \int    \frac{1}{\sqrt{ \tau^2+s_i^2 }}   \phi\left(
\frac{  r - \theta }{\sqrt{ \tau^2+ s^2}} \right)  dF(s) dr  \\
& =&  \int \Phi\left( \frac{  t - \theta }{\sqrt{ \tau^2+ s^2}} \right)  dF(s),
\eas
where the second equality follows from Equation \eqref{lemma-prob-eq2}.
Based on the  MLE $\widehat F(x)$ of $F(x)$,
we immediately obtain the  MLE of $G(t)$,
\bas
\widehat G( t)
& =&   \int \Phi\left( \frac{  t - \widehat \theta }{\sqrt{ \widehat \tau^2+ s^2}} \right)  d\widehat F(s)
= \sum_{i=1}^n \widehat p_i \Phi\left( \frac{  t - \widehat \theta }{\sqrt{ \widehat \tau^2+ s_i^2}} \right).
\eas
By Theorem \ref{asy-full-likelihood},
$\widehat \bgamma$  and $\widehat \alpha$ are  consistent,
therefore   $\widehat F$ and $\widehat G$ are also consistent estimators
of $F$ and $G$, respectively.

\subsection{Comparison with conditional likelihood }

In the literature,  the conditional likelihood  $\ell_c(\bgamma)$ is usually used to estimate
the underlying parameters for the Copas-like model.
Let $\tilde \bgamma = \arg\max_{\bgamma}\ell_c(\bgamma)$
be the  conditional  MLE of $\bgamma$.
Then  $N$ can be estimate by the inverse probability weighting estimator
or the  MLE
\bas
\tilde N   &=&  \sum_{i=1}^n  \frac{1}{ f_3(  s_i; \tilde \bgamma_{12}) }
= \sum_{i=1}^n  \frac{1}{ \Phi(\tilde \gamma_1 +\tilde  \gamma_2/s_i)}.
\eas

\begin{theorem}
	\label{asy-conditional-mle}
	Assume Condition C1  in the supplementary materials,
	$\rho_0\neq 0$,   and that $\bV_c$ is positive definite.
	Then as $N_0\rightarrow\infty$,
	(i) $N_0^{1/2}(\tilde \bgamma -\bgamma_0) \convergeto N(\bzero, \bV_c^{-1})$, and
	(ii) $N_0^{1/2}(\tilde N /N_0 - 1) \convergeto N(0, \sigma_c^2)$,
	where   $\sigma_c^2 =  \varphi_1-1 +   \bvarphi_{2}^\T
	\bE_{12} \bV_c^{-1} \bE_{12}^\T \bvarphi_{2}$.
\end{theorem}

We may wonder whether the proposed MLEs have efficiency gain over
the  conditional  MLEs  in terms of asymptotical variance.
Unfortunately  the answer is negative.

\begin{proposition}
	\label{efficiency-sigma}
	With the symbols used in  Theorems
	\ref{asy-full-likelihood} and \ref{asy-conditional-mle},
	$\sigma^2 =  \sigma_c^2$ and $\bV = \bV_c $.
\end{proposition}

This proposition indicates that to estimate  $N$ and $\bgamma$,
the conditional MLEs and proposed MLEs
have the same asymptotic  normal distribution.
The proposed full likelihood estimation procedure
has no efficiency improvement over
the conditional likelihood estimation procedure.
Even so, the proposed full likelihood method still has
several advantages over the conditional likelihood method.
First,  although the resulting point estimators are asymptotical equivalent,
the interval estimators based on these two methods generally have
quite different finite-sample performances.
Our simulation results indicate that
the proposed likelihood ratio interval
usually has better coverage accuracy than
the conditional-likelihood-based Wald interval.
Second, the proposed likelihood ratio interval is free from variance estimation,
which however is inevitable for the conditional-likelihood-based Wald interval.
Third, the Wald interval estimators
may have so small lower bounds that are even less than
the number of studies observed in the meta analysis, which
is clearly unreasonable. By contrast,
the proposed likelihood ratio interval estimators
never suffer from such an embarassment.
Finally,  under the conditional likelihood method,
similar to $\tilde N$,  $F(x)$ and $G(t)$ are also estimated
by their inverse probability weighting estimators
\bas
\tilde    F(s)
&=&
\frac{1}{\tilde N} \sum_{i=1}^n  \{ \Phi(\tilde \gamma_1+\tilde \gamma_2/s_i) \}^{-1} I( s_i \leq s)
= \frac{ \sum_{i=1}^n  \{ \Phi(\tilde \gamma_1+\tilde \gamma_2/s_i) \}^{-1} I( s_i \leq s)   }{
	\sum_{j=1}^n  \{ \Phi(\tilde \gamma_1+\tilde \gamma_2/s_j) \}^{-1} }, \\
\tilde    G(t)
&=& \frac{ \sum_{i=1}^n  \{ \Phi(\tilde \gamma_1+\tilde \gamma_2/s_i) \}^{-1} \Phi\{ ( t - \tilde\theta)/\sqrt{
		\tilde \tau^2+ s_i^2} \}  }{
	\sum_{j=1}^n   \{ \Phi(\tilde \gamma_1+\tilde \gamma_2/s_j) \}^{-1}
}.
\eas
Because of inverse  probability weighting,  it is well known that
the numerical performance of inverse probability weighting estimators can be quite unstable
when some of $\Phi(\tilde \gamma_1+\tilde \gamma_2/s_i)$  are close to 0.
The use of EL in our full likelihood estimator considerably mitigates this embarassment.
The maximization of the empirical likelihood $\prod_{i=1}^n p_i$
greatly prevents the occurrence of extreme small weights, and thus
leads to more stable numerical performances of $\widehat F(s)$.
This numerical advantage of  the proposed MLE over the inverse probability weighting estimator
was also noticed by  \cite{Han2014} .

\subsection{Calculation of MLEs}

Maximizing the full likelihood $\ell(N, \alpha, \bgamma)$
is computationally challenging because  the function
$\ell(N, \alpha, \bgamma)$
takes maximum over a very flat plateau
and its maximization often produces non-convergence results.
If $\bgamma_{12}$ is fixed, then the maximization
can be stably obtained.  This phenomenon has been acknowledged by \cite{Copas2001}
and also observed by \cite{Ning2017}  in maximizing the conditional likelihood $\ell_c$.
To overcome this problem,  \cite{Ning2017}  proposed an expectation-maximization algorithm after
recasting the bias sampling issue as a missing data problem.
Their simulation studies indicate that the
expectation-maximization algorithm usually produces stable estimates for $\bgamma$.
Let $\tilde \bgamma = (\tilde \bgamma_{12}, \tilde \rho, \tilde \tau, \tilde \theta)$
be the conditional estimate of $\bgamma$ calculated
by \cite{Ning2017} 's expectation-maximization algorithm.
Since direct maximization with respect to $\bgamma_{12}$ is very unstable,
we propose to maximize our full log-likelihood $\ell(N, \alpha, \bgamma)$
by fixing $\bgamma_{12} = \tilde \bgamma_{12}$.

For ease of exposition, we first re-express the profile empirical log-likelihood  as
\bas
\ell(N, \alpha, \bgamma)=h_1(N,\alpha)+h_2(\bgamma)
+  \min_{\lambda} h_3(\bgamma_{12},\alpha,\lambda),
\eas
where
\bas
h_1(N,\alpha)
&=& \log { N \choose  n} + (N-n) \log(1-\alpha),  \\
h_2(\bgamma)
&=&
\sum_{i=1}^n \left[ \log \{ \Phi(v_i) \}
-\frac{1}{2} \log(\tau^2 + s_i^{ 2} )
-  \frac{ (\theta_i - \theta)^2 }{2(\tau^2 + s_i^{ 2})} \right],  \\
h_3(\alpha, \bgamma_{12},\lambda)
&=&
-\sum_{i=1}^n \log[ 1+\lambda\{ \Phi(\gamma_1+\gamma_2/s_i ) - \alpha \} ].
\eas
To avoid the non-definition problem in EL,
we adopt  \cite{Owen1990}'s calculation strategy  and replace
the $\log(\cdot)$ function  in $h_3$   by
\begin{equation}
\log_*(z)=\left\{ \begin{array}{c@{,}c}
\log(z)  & \quad z>1/c_n \\
-\log(c_n)-1.5+ 2z c_n - 0.5 z^2 c_n^2    & \quad z\leq 1/c_n
\end{array}\right.,
\end{equation}
where $c_n $  is a pre-specified large number and is usually chosen to be $n$.

We propose to maximize $\ell(N, \alpha, \bgamma)$ via the following algorithm:
\begin{description}
	
	\item[Step 1] 	Calculate   $\tilde h_1(\alpha) = \max_N h_1(N,\alpha)$, and
	$\tilde h_3(\alpha, \bgamma_{12}) = \min_\lambda h_3(\alpha, \bgamma_{12}, \lambda)$.
	\item[Step 2]	Let $h_{23}(\alpha, \bgamma)=  h_2(\bgamma) + \tilde h_3(\alpha, \bgamma_{12})$.
	Calculate $\tilde h_{23}(\alpha) = \max_{\bgamma} h_{23}(\alpha, \bgamma)$.
	\item[Step 3]	Let $h_{123}(\alpha)=\tilde h_1(\alpha)+ \tilde h_{23}(\alpha)$.
	Calculate $\max_{\alpha} h_{123}(\alpha)$
	and the maximizer $\hat \alpha$.
	\item[Step 4]	Calculate $\widehat N = \arg\max_N\ h_1(N, \widehat\alpha)$
	and   $\widehat \bgamma = \arg\max_{\bgamma} h_{23}(\bgamma,\widehat\alpha)$.
\end{description}
As $\bgamma$ is a 5-variate vector, we   implement
the optimizations  with respect to $\bgamma$  by the R command
{\tt nlminb}.
All the other optimizations in the above algorithm are
with respect to a scalar variable and can be quickly solved
by the R command {\tt optimize}.

\section{Simulations}

\subsection{Simulation settings}
We carry out simulations to investigate the finite-sample performance
of the proposed full likelihood method (Full Likelihood or FL)
and compare  it with the  conditional likelihood method
(Conditional Likelihood or CL)
implemented by  \cite{Ning2017} 's expectation-maximization algorithm.
Consistent estimators are needed for $\sigma_c^2$ and $\bV_c$
when we apply the conditional likelihood method to construct Wald type intervals for $N$ and $\bgamma$.
Following \cite{Ning2017} ,  we adopt the estimation procedure
by \cite{Louis1982} together with  \cite{Ning2017} 's expectation-maximization algorithm
to obtain consistent variance estimators for $\tilde N$ and $\tilde \bgamma$.

We generate study-specific variance $s_i^{*2}$ from the square of a normal
random variable $N(0.25,0.5)$, and generate  $(\epsilon_i,\delta_i)$'s  from
a bivariate standard normal distribution with correlation $\rho_0$.
Given $N_0$ and $ \bgamma_0$, we calculate $\theta_i^*$ and $Z_i$
from models \eqref{rem}  and \eqref{lv} with $\bgamma_0$ in place of $\bgamma$.
The $(\theta_i^*, s_i^*)$'s with $Z_i>0$ constitute a simulated sample.
We  consider two choices of $N_0$, $50$ and $100$,
and two choices of $\bgamma_{12, 0}$,    $ (-0.6,0.8)$ and $ (-1, 0.6)$.
As $\bgamma_{12, 0}$ changes from  $ (-0.6,0.8)$ to $ (-1, 0.6)$,
the publishing rate decreases from 80\% to 64\%,
publication bias getting more and more severe.
To examine the effects of effect size,
publication bias  and  heterogeneity on
the performances of the full likelihood and conditional likelihood methods,
we consider two scenarios for $(\theta_0, \tau_0, \rho_0)$:
(1)   $\theta_0=0.4$, $\tau_0 =0.5$, $\rho_0 = 0.2, 0.8$,
and (2)  $\theta_0=0.2$,  $\rho_0 =0.2$, $\tau_0 = 0.5, 1$.

For each parameter combination, we generate 1000 simulation samples,
and calculate the full likelihood and conditional likelihood point estimates
for $N, \theta$,  and $\tau$ based on each sample.
The simulated bias (BIAS), standard deviation (SD) and root mean square error (RMSE)
of these two type estimators are then obtained.
We also  calculate the simulated coverage probabilities of
the proposed likelihood ratio confidence intervals
and the conditional-likelihood-based Wald confidence intervals
for the three parameters at the 95\% confidence level.
These simulation results
are tabulated in Tables \ref{simu-table1}  and \ref{simu-table2},
corresponding to scenarios (1) and (2), respectively.

\begin{table}
	
	\caption{Simulation results for scenario (1) with $\tau_0=0.5$ and different choices of $\rho$.
		All numbers have been multiplied by 100. PAR: parameter; SD: standard deviation;
		BIAS: bias;RMSE: root mean square error;CP:  coverage probability at 95\% confidence level.}
	\label{simu-table1}
	\begin{center}
		\small
		\begin{tabular}{ccccc cccccccc}
			\hline
			& & &  & \multicolumn{4}{c}{ Conditional Likelihood  } & & \multicolumn{4}{c}{Full Likelihood} \\ \cline{5-8}\cline{10-13}
			$\gamma_{12,0}$   &$\rho_0$& $N_0$ & PAR & BIAS & SD & RMSE & CP & &BIAS &SD & RMSE & CP\\ \hline
			$(-0.6, 0.8)$& 0.2 & 50 & $\theta$ &  -3.44&   11.93&   12.42&  95.3&&    0.22&    11.58&   11.59&  94.3  \\
			&     &    & $\tau$   &  -1.81&   14.23&   14.34&  89.9&&   -0.96&     8.88&    8.93&  94.0   \\
			&     &    & $\rho$   &  21.09&   47.95&   52.38&  100 &&   -2.02&    52.49&   52.53&  92.0   \\
			&     &    & $N$      &  23.11&  428.47&  429.09&  89.8&&  -44.31&   428.66&  430.95&  93.4 \\ 
			&     & 100& $\theta$ &  -3.39&    7.85&    8.55&  95.3&&   -0.07&     8.12&    8.12&  95.1   \\
			&     &    & $\tau$   &  -2.36&   16.02&   16.19&  90.3&&   -0.91&     6.14&    6.20&  94.5   \\
			&     &    & $\rho$   &  20.51&   37.87&   43.06&  100 &&   -0.85&    36.25&   36.26&  93.6   \\
			&     &    & $N$      &  28.56&  609.81&  610.48&  90.0&&  -38.71&   609.98&  611.21&  94.3   \\ 
			& 0.8 & 50 & $\theta$ &    3.08&    13.64&    13.98&  93.5&&   -0.35&    11.69&    11.70&  92.4   \\
			&     &    & $\tau$   &   -3.17&    13.85&    14.21&  96.3&&   -2.32&     8.34&     8.65&  94.2    \\
			&     &    & $\rho$   &  -15.37&    57.47&    59.49&  100 &&   -8.49&    49.56&    50.37&  96.3   \\
			&     &    & $N$      &  207.55&   438.85&   485.45&  87.1&&  138.78&   439.21&   460.62&  95.6 \\ 
			&     & 100& $\theta$ &    2.01&     8.93&     9.15&  95.1&&   -0.48&     8.99&     9.01&  93.4   \\
			&     &    & $\tau$   &   -1.35&     9.39&     9.49&  95.3&&   -1.50&     6.31&     6.48&  95.9  \\
			&     &    & $\rho$   &  -12.00&    40.66&    42.40&  100 &&   -4.34&    35.83&    36.09&  94.8   \\
			&     &    & $N$      &  464.97&   650.34&   799.46&  91.3&&  396.54&   650.59&   761.92&  94.0   \\ \hline
			$(-1 ,0.6)$  & 0.2 & 50 & $\theta$ &  -4.69&   14.64&   15.36&  93.8&&  -0.004&    13.38&   13.37&  92.6  \\
			&     &    & $\tau$   &  -1.02&   13.79&   13.82&  92.3&&   -2.40&     9.33&    9.64&  94.1   \\
			&     &    & $\rho$   &  13.91&   49.36&   51.28&  100 &&    2.11&    42.98&   43.03&  93.5    \\
			&     &    & $N$      &  99.71&  769.12&  775.55&  89.9&&   18.24&   770.62&  770.84&  92.8 \\ 
			&     & 100& $\theta$ &  -4.17&   10.56&   11.35&  92.5&&   -0.18&     9.12&    9.12&  94.3   \\
			&     &    & $\tau$   &   0.15&   11.02&   11.02&  91.7&&    0.89&     6.59&    6.65&  93.5   \\
			&     &    & $\rho$   &  14.13&   38.18&   40.71&  99.9&&    1.01&    30.35&   30.37&  93.3 \\
			&     &    & $N$      & 255.43& 1003.34& 1035.35&  90.1&&  174.44&  1004.56& 1019.59&  94.8\\ 
			& 0.8 & 50 & $\theta$ &  -16.75&    15.38&    22.74&  92.0&&   -0.98&    13.77&    13.81&  91.5   \\
			&     &    & $\tau$   &   -2.53&    10.76&    11.05&  94.3&&   -1.99&     9.29&     9.50&  94.0   \\
			&     &    & $\rho$   &   38.75&    82.42&    91.08&  100 &&   23.84&    34.84&    42.22&  96.7    \\
			&     &    & $N$      &  308.42&   830.88&   886.28&  88.1&&  229.04&   831.31&   862.29&  91.1  \\ 
			&     & 100& $\theta$ &   -8.75&    11.80&    14.69&  92.3&&   -1.25&     8.85&     8.94&  92.7    \\
			&     &    & $\tau$   &  -14.32&     8.07&    16.45&  94.7&&   -0.82&     6.19&     6.24&  94.4    \\
			&     &    & $\rho$   &   -3.94&    68.64&    68.76&  100 &&   -2.72&    28.50&    28.64&  95.2  \\
			&     &    & $N$      &  549.06&  1248.90&  1364.27&  92.7&&  469.78&  1249.01&  1334.43&  93.3  \\
			\hline
		\end{tabular}
	\end{center}
\end{table}

\begin{table}
	\caption{Simulation results  for scenario (2) with $\rho_0= 0.2$  and different choices of $\tau$.
		All numbers have been multiplied by 100. PAR: parameter; SD: standard deviation;
		BIAS: bias;RMSE: root mean square error;CP:  coverage probability at 95\% confidence level.}
	\label{simu-table2}
	\begin{center}
		\small
		\begin{tabular}{ccccccccccccccc}
			\hline
			&  & & & \multicolumn{4}{c}{ Conditional Likelihood  } & & \multicolumn{4}{c}{Full Likelihood} \\
			\cline{5-8}\cline{10-13}
			$\gamma_{12, 0}$   &$\tau_0$& $N_0$ & PAR  & BIAS&  SD & RMSE & CP & &BIAS &SD & RMSE & CP\\ \hline
			$( -0.6, 0.8)$ & 0.5 & 50 & $\theta$ &   -3.24&   11.31&    11.76&  95.3&&   0.006&    11.77&    11.77&  93.7  \\
			&     &    & $\tau$   &    1.96&   11.88&    12.04&  93.4&&   -2.02&     8.61&     8.85&  95.4  \\
			&     &    & $\rho$   &   18.34&   52.73&    55.85&  100 &&   -2.22&    50.08&    50.13&  92.8  \\
			&     &    & $N$      &  161.76&  442.90&   471.52&  87.2&&   93.72&   443.32&   453.12&  94.3  \\
			&     & 100& $\theta$ &   -3.87&    7.81&     8.71&  94.7&&    0.29&     8.06&     8.06&  94.3  \\
			&     &    & $\tau$   &    0.27&    8.82&     8.82&  91.8&&   -0.39&     6.23&     6.25&  94.8  \\
			&     &    & $\rho$   &   18.54&   30.77&    35.93&  100 &&   -2.51&    35.50&    35.59&  92.6  \\
			&     &    & $N$      &  284.71&  596.57&   661.03&  87.3&&  217.02&   596.78&   635.01&  95.0   \\
			&  1  & 50 & $\theta$ &  -10.33&   21.92&    24.23&  93.4&&    0.59&    20.69&    20.70&  94.5  \\
			&     &    & $\tau$   &    0.26&   17.55&    17.55&  92.2&&    2.39&    14.00&    14.20&  93.9  \\
			&     &    & $\rho$   &   37.32&   91.12&    98.46&  100 &&   -2.39&    61.37&    61.42&  93.8  \\
			&     &    & $N$      &  153.37&  449.37&   474.83&  87.4&&   84.35&   449.61&   457.45&  94.9  \\
			&     & 100& $\theta$ &   -2.05&   12.59&    12.76&  97.7&&   0.006&    10.44&    10.44&  94.6  \\
			&     &    & $\tau$   &   -0.43&   10.11&    10.10&  94.7&&   -0.39&     9.88&     9.85&  94.8  \\
			&     &    & $\rho$   &    9.05&   52.60&    53.37&  100 &&   -0.91&    48.30&    48.30&  93.6 \\
			&     &    & $N$      &  134.07&  590.23&   605.26&  90.3&&   66.17&   590.48&   594.18&  94.8    \\\hline
			$(-1, 0.6)$  & 0.5 & 50 & $\theta$ &   -4.52&   14.65&    15.33&  91.7&&   -0.42&    12.87&    12.88&  93.7  \\
			&     &    & $\tau$   &   -1.79&   18.14&    18.23&  90.5&&   -2.19&     9.40&     9.65&  93.2  \\
			&     &    & $\rho$   &   16.47&   61.13&    63.31&  99.9&&    2.80&    43.07&    43.16&  92.4  \\
			&     &    & $N$      &   33.16&  703.52&   704.30&  86.7&&  -47.01&   704.76&   706.31&  93.3  \\
			&     & 100& $\theta$ &   -5.66&   12.09&    13.35&  92.6&&    1.07&     8.65&     8.72&  95.2  \\
			&     &    & $\tau$   &    1.54&   11.32&    11.42&  91.9&&   -0.96&     6.11&     6.18&  95.5  \\
			&     &    & $\rho$   &   18.41&   48.78&    52.14&  99.7&&    2.42&    35.34&    35.43&  91.9 \\
			&     &    & $N$      &  292.59& 1027.42&  1068.27&  89.7&&  210.83&  1028.20&  1049.59&  95.0 \\
			&  1  & 50 & $\theta$ &   -5.72&   24.20&    24.87&  95.0&&    1.91&    22.33&    22.41&  94.0  \\
			&     &    & $\tau$   &   -0.58&   18.19&    18.20&  93.1&&   -2.37&    14.80&    14.91&  94.6  \\
			&     &    & $\rho$   &   15.83&   74.37&    76.04&  100 &&   -3.42&    55.56&    55.66&  93.9  \\
			&     &    & $N$      &   46.74&  685.31&   686.90&  89.6&&  -33.69&   686.24&   687.01&  94.7  \\
			&     & 100& $\theta$ &   -7.08&   15.94&    17.45&  94.9&&   -0.28&    16.28&    16.28&  95.0  \\
			&     &    & $\tau$   &    0.08&   12.89&    12.89&  92.7&&   -1.41&    10.36&    10.46&  95.3   \\
			&     &    & $\rho$   &   15.54&   48.86&    51.27&  100 &&   -1.49&    40.88&    40.91&  93.3   \\
			&     &    & $N$      &  113.87& 1041.61&  1047.81&  91.8&&   33.62&  1042.29&  1042.83&  93.7  \\
			\hline
		\end{tabular}
	\end{center}
\end{table}

\subsection{Simulation results}
We first  examine the results on  point estimation.
The full likelihood estimators of all the four parameters  $\theta, \tau, \rho$ and $N$
have obviously smaller RMSEs than the conditional likelihood estimators in almost
all scenarios (58 out of 64).  In the rest 8 scenarios,
although the full likelihood estimators do not win, their performances
are  nearly the same as the conditional likelihood estimators in terms of RMSE.
These observations indicate  that the proposed full likelihood  method has clear
advantages over the traditional conditional likelihood method.
Meanwhile the efficiency gain of the full likelihood estimators
increases  as the effect size $\theta$ increases
(from 0.2 to 0.4) or  the publication bias
becomes more severe (as $\rho$ increases from 0.2 to 0.8).
When estimating $\theta$,  $N$ and $\rho$,
the full likelihood estimators usually have smaller absolute bias
although their standard deviations are very close
to those of the conditional likelihood estimators.
In particular for $\rho$, the full likelihood estimator corrects
most part of  the conditional likelihood estimator's bias.
While when estimating $\tau$,  the proposed full likelihood estimator
usually has smaller standard deviation,
although the full likelihood and conditional likelihood estimators have very close biases.
In all cases, as $N_0$ increases from 50 to 100,
the values of BIAS, SD and RMSE  decrease as expected
when the parameters are $\theta$, $\tau$ and $\rho$.
When $N$ is of interest, in  almost all cases,
the relative BIAS,  relative  SD and relative  RMSEs of both estimators
decrease as $N_0$ increases.

For interval estimation, the proposed likelihood ratio interval
or the full likelihood interval  for the four parameters
almost always has more accurate and more reliable coverage probabilities than
the conditional likelihood-based Wald  interval  or the conditional likelihood interval.
For example,  in the cases of $\bgamma_{12, 0}=(-0.6, 0.8)^\T $, $\theta=0.4$, and $\rho_0=0.8$,
the coverage probabilities of the conditional likelihood interval for $N$
are only 87.1\%  and 91.3\%  when the true value of $N$ is 50 and 100, respectively.
By contrast, the corresponding numbers of the full likelihood interval
are 95.6\% and 94.0\%, respectively, which are much more desirable.
The advantage of the full likelihood interval  is more evident in the estimation of $\rho$.
The coverage probabilities of the full likelihood intervals
varies from 92.1\%  to 95.3\%, which are reasonable.
However,  the coverage probabilities
of the conditional likelihood interval for $\rho$  are almost always 100\%, indicating that
this interval is  too wide to be practically useful.
In addition, the full likelihood interval exhibits  more robust
performance than the conditional likelihood interval in terms of coverage probability
as the simulation setting varies.

\subsection{Comparison in QQ-plots}
\begin{figure}[b]
	\centering		
	\includegraphics[width=2in]{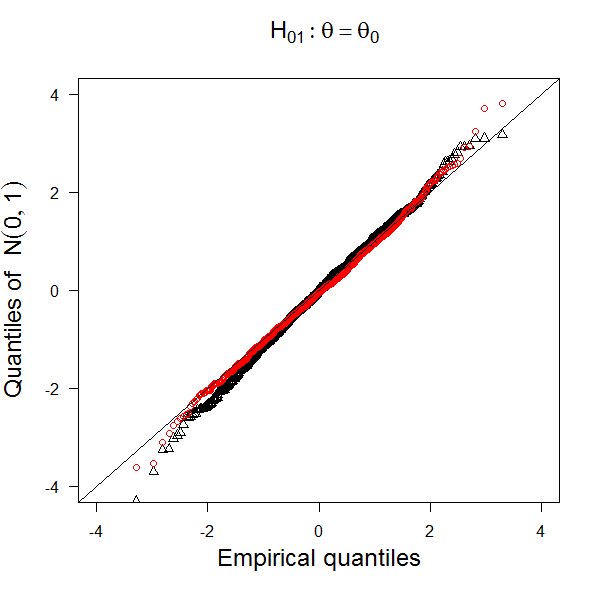}
	\includegraphics[width=2in]{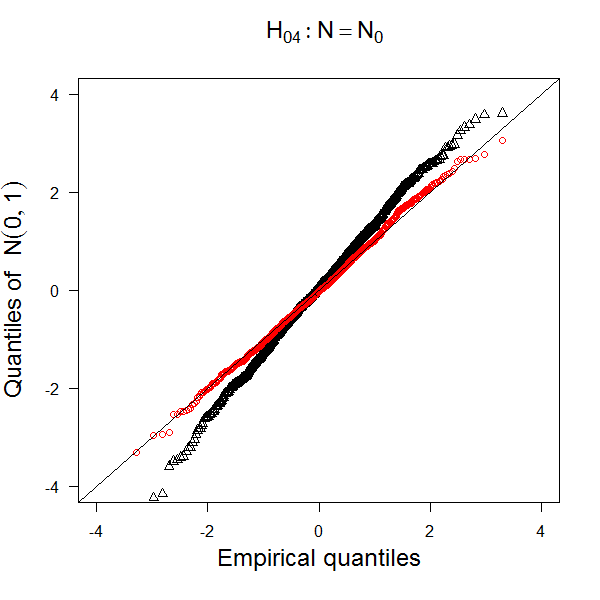}
	\caption{
		QQ-plots of the Wald statistic (triangle) and the sign-root of the proposed likelihood ratio
		statistic (circle) versus N(0, 1).
		The  test statistics are calculated based on
		1000 samples from the Copas-like selection model with
		$N_0 =50$, $\theta_0 =0.2$, $\tau_0 = 1$, $\bgamma_{12, 0}=(-0.6, 0.8)$,
		and $\rho_0 = 0.2$.}
	\label{plot-qq-lrt-wald}
\end{figure}

To get more insights about the better performance of the full likelihood intervals over the conditional likelihood intervals,
we consider the following four hypothesis testing problems:
$H_{01}: \theta=\theta_0$, $H_{02}: \tau=\tau_0$, $H_{30}: \rho=\rho_0$,
and $H_{04}: N = N_0$.
We  generate  1000 samples  from the Copas-like model with
$\theta_0=0.2, \tau_0=1, \rho_0=0.2$,
$\bgamma_{12,0} =(-0.6, 0.8)$   and $N_0=50$.
For each of the four hypothesis testing problem,
1000 likelihood ratio statistics  and
Wald test statistics   were calculated.
The  qq-plots of the sign-roots  of the 1000 likelihood ratio statistics  and   Wald test statistics  versus
the standard normal quantiles are displayed in  Figure  \ref{plot-qq-lrt-wald}.

Clearly, the qq-plots of the sign-roots of the likelihood ratio statistics
are all quite close to the identity line
for all the four hypothesis testing problems.
This implies that N(0, 1)  and $\chi_1^2$
are desirable approximates to the finite-sample distributions of
the sign-root  and itself of the likelihood ratio statistic.
It also explains why the full likelihood intervals for the four parameters
always have very nice coverage accuracy.
By contrast, the qq-plots of the Wald test statistic
are not that close to the identity line.
The departure becomes larger and larger
from  $H_{01}$ to $H_{02}$  to $H_{04}$,
which explains the poorer and poorer coverage probabilities
(from 93.4\% to 92.2\% to 87.4\%) of the conditional likelihood interval.
For $H_{03}$,  the Wald statistic has  much larger lower quantiles
and  much smaller upper quantiles  compared with the standard normal.
This makes the coverage probability of
the resulting conditional likelihood interval  unacceptably large
since its construction is based on the Wald  statistic
calibrated by the standard normal.

In summary, the limiting $\chi^2$ distribution always  approximates much  better to
the finite-sample distribution of
the likelihood ratio statistic
than the standard normal to that of
the Wald statistic.
This makes the resulting full likelihood intervals
always have more, sometimes far more, accurate
coverage probabilities than the  conditional likelihood intervals.

\section{Premature birth  data}

For further comparison of the full likelihood and conditional likelihood methods, we apply them
to a meta analysis, in both of which $\theta$ stands for log-odds ratio.
The data for meta analysis comes from  \cite{Copas2004}.
It consists of  the results of 14 randomized clinical trials
concerning the use of prophylactic corticosteroids  in cases of premature birth.
The treatment is administered to the mother in order to improve
the chance of the infant's survival if a birth is anticipated to be premature.
Table \ref{table-realdata} reports the analysis results of
the  full likelihood and conditional likelihood methods.

\begin{figure}[b]
	\centering
	\includegraphics[width=2.5in,height=2.5in]{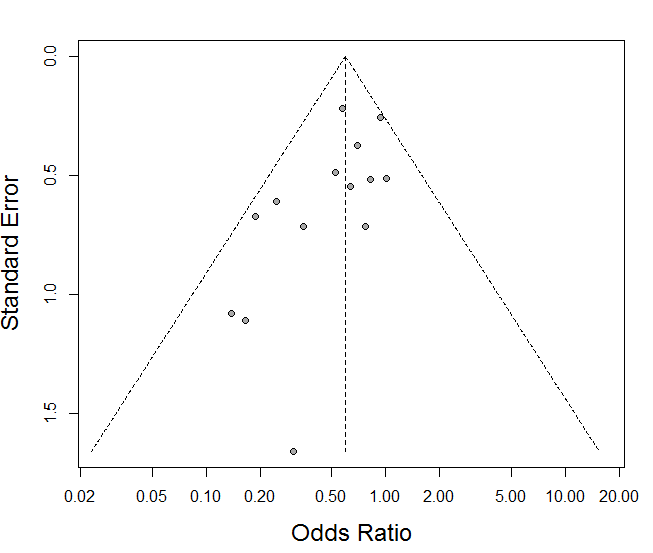}
	\caption{
		Funnel plot  of   the Premature birth data.}
	\label{fig:funnel}
\end{figure}

The funnel plot of this data, shown in the right panel of
Figure \ref{fig:funnel}, looks quite asymmetric,
indicating that publication bias does exists.
This also coincides with the observation from
our likelihood ratio confidence interval for $\rho$ at the 95\% level,
$[-1, -0.025]$, which excludes 0.
However the Wald confidence based on the conditional likelihood
is $[-2.543, 0.586]$, seemingly supporting $\rho=0$,
which  is clearly unreliable.
In addition,  both confidence intervals for $\tau$
include zero, which provides certain evidence for
the non-existence of  between-study heterogeneity.

The parameter  $\theta$ denotes the underlying log-odds ratio comparing the probability
of death in the treated group with that for a parallel sample of controls.
Both the full likelihood and conditional likelihood methods give the same point estimate, -0.476,
for  $\theta$ with similar intervals [-0.760,-0.244]
and [-0.662, -0.289].
The log-odds ratio estimate -0.476 implies that
the use of prophylactic corticosteroids  in cases of premature birth
can reduce morality by  as large as 28\%.
Hence our meta analysis provides strong support for
the use of prophylactic corticosteroids in cases of premature birth.

For point estimation of $N$,
the full likelihood and conditional likelihood estimates are 23 and 16, respectively.
However for interval estimation,
again the lower bound (7) of the Wald interval  is
less than the number of observed studies (13),
while  that  of the likelihood ratio  interval  is  13,
which makes more sense.

\begin{figure}[b]
	\centering
	\includegraphics[width=2in]{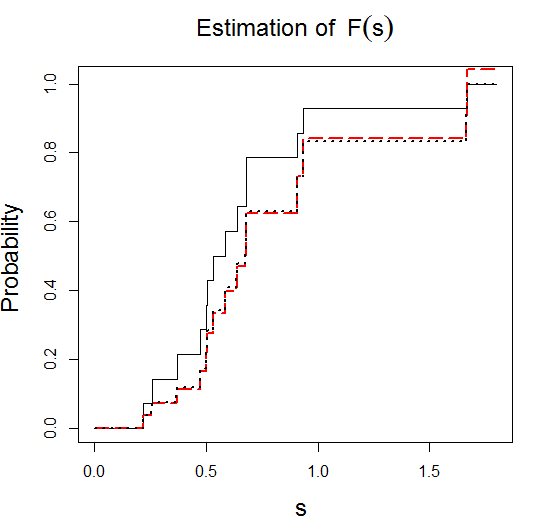}
	\includegraphics[width=2in]{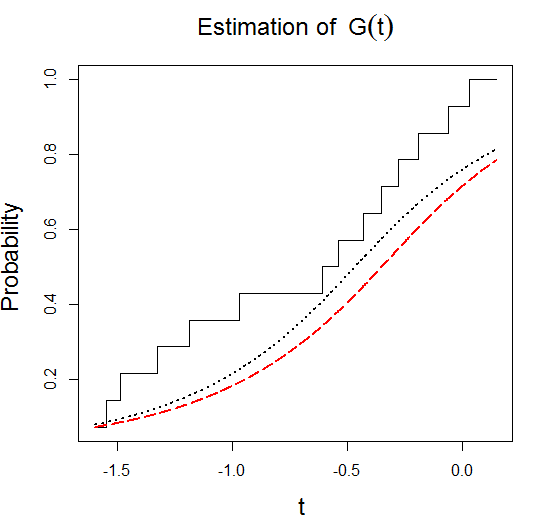}
	\caption{
		Display of empirical distribution, the FL distribution estimate  (dashed line)
		and the CL distribution estimate  (dotted line)
		for   the Premature birth data.}
	\label{distribution}
\end{figure}

In the presence of publication bias,
the published studies, $\{(\theta_i, s_i): i=1,2, \ldots, n \}$,
constitute a biased sample of all studies;
the empirical distributions of $s_i^*$'s and $\theta_i^*$'s respectively
are inconsistent estimators of  the underlying population distributions,
$F(s)$ and $G(t)$.
Theoretically both the proposed full likelihood and the conditional likelihood methods
can  correct publication bias  and
consistently estimate $F(s)$ and $G(t)$.
In Section 2, we have presented  the  MLEs $\widehat F(s)$ and
$ \widehat G(t)$ based on the full likelihood method,
and the inverse probability weighting estimators $ \tilde F(s)$
and $\tilde G(t)$ based on the conditional likelihood method.
Figure ~\ref{distribution} displays
the empirical distributions, the full likelihood and conditional likelihood estimators
for both  $F(s)$ and $G(t)$ based on the premature birth data.
As publication bias very likely exists in this data, we observe that
the full likelihood and conditional likelihood estimates
are away from the empirical distribution.

\begin{table}
	\caption{Meta analysis results of the lung cancer data and the premature birth  data.
		Est: estimate value; CI: 95\% confidence intervals ;$\theta$:log odds ratio; $\tau$: between-study heterogeneity}
	\begin{center}
		\begin{tabular}{cccccc}
			\hline
			& \multicolumn{2}{c}{ Conditional Likelihood  } & & \multicolumn{2}{c}{Full Likelihood} \\
			\cline{2-3}\cline{5-6}
			& Est & CI  &  & Est &CI  \\ \hline
			& \multicolumn{4}{c}{Premature birth  data}  \\
			$\theta$        & -0.476    & [-0.662,-0.289] &  & -0.476   & [-0.760,-0.244]     \\
			$\exp(\theta)$  & 0.621     & [0.516,0.748]   &  &  0.621   & [0.468,0.784]       \\
			$\tau$          & 0         & [-0.240,0.240]  &  &  0       & [-0.201,0.484]       \\
			$\rho$          & -0.977    & [-2.543,0.586]  &  & -0.837   & [-1.000,-0.025]      \\
			$N$             & 23        & [7,  39]        &  & 16       & [13,  20]            \\  \hline
		\end{tabular}
	\end{center}
	\label{table-realdata}
\end{table}

\section{Discussion}

To correct publication bias in meta analysis,
we propose a full likelihood semi-parametric approach under the Copas-like
selection model of \cite{Ning2017} .
We have demonstrated the advantages of the proposed full likelihood method
over the commonly-used conditional likelihood and Wald-type method
by theoretical and numerical studies.
We show that the full MLEs
have smaller mean squared errors than
the conditional-likelihood-based estimators.
The full likelihood ratio confidence intervals for
the effect size and the total number of studies
have more accurate coverage probabilities than
the  Wald intervals under the conditional likelihood.

A key issue in the implementation of our method is that
the maximization of the full likelihood is numerically very difficult,
as the data contains little information about
$\gamma_1$ and $\gamma_2$.
This problem exists also in the maximization of the condition likelihood
as pointed out by  \cite{Ning2017}.
To then end,  we fix $\gamma_1$ and $\gamma_2$ to be their
maximum conditional likelihood estimates that are calculated
with \cite{Ning2017}'s expectation-maximization algorithm.
In doing so,  the resulting parameter estimates are
in essence different from  the true maximum likelihood estimates.
Similar to the conditional likelihood (see  \cite{Ning2017}),
the full likelihood function seems to be
a very flat plateau around its maximum.
This also implies that the replacement of the true MLEs  $\gamma_1$ and $\gamma_2$
with their conditional MLEs does not lead much change in
the full likelihood ratio test statistics
with respect to parameters other than  $\gamma_1$ and $\gamma_2$.

\begin{acknowledgements}
 Dr.~Liu's research  was supported by the National Natural Science
Foundation of China (11771144, 11971300, 11871287),  the State Key Program of the National
Natural Science Foundation of China (71931004),
the Natural Science Foundation of Shanghai  (19ZR1420900, 17ZR1409000),
the development fund for Shanghai talents , the 111 project (B14019), and
the Fundamental Research Funds for the Central Universities.
Dr.~Li was supported in part by  the Natural Sciences and
Engineering Research Council of Canada grant number  RGPIN-2015-06592.
\end{acknowledgements}

\section*{Supplementary material}

The Supplementary Material contains detailed proofs for Lemma 1, Theorems 1 and 2, and
Proposition 1.


 \bibliographystyle{plainnat}
\bibliography{file}

%
%
%
%
%
	


\end{document}